\documentclass{emulateapj}

\shorttitle{Hybrid stars in the light of the massive pulsar PSR J1614-2230}
\shortauthors{C. H. Lenzi and G. Lugones}

\begin{document}

\title{Hybrid stars in the light of the massive pulsar PSR J1614-2230}
\author{C. H. Lenzi and G. Lugones}
\affiliation{Universidade Federal do ABC, Rua Santa Ad\'elia 166, Santo Andr\'e, SP, 09210-170, Brazil.}
\email{cesar.lenzi@ufabc.edu.br} 
\email{german.lugones@ufabc.edu.br}

\begin{abstract}
We perform a systematic study of hybrid star configurations using several parametrizations of a relativistic mean-field hadronic EoS and the NJL model for three-flavor quark matter. For the hadronic phase we use the stiff GM1 and TM1 parametrizations, as well as the very stiff NL3 model. In the  NJL Lagrangian we include scalar, vector and 't Hooft interactions. The vector coupling constant $g_v$ is treated as a free parameter.  
We also consider that there is a split between the deconfinement and the chiral phase transitions which is controlled by changing the conventional value of the vacuum pressure $- \Omega_0$ in the NJL thermodynamic potential by $- (\Omega_0 + \delta \Omega_0)$, being  $\delta \Omega_0$  a free parameter.
We find that, as we increase the value of $\delta \Omega_0$,  hybrid stars have a larger maximum mass but are less stable, i.e. hybrid configurations are stable within a smaller range of central densities. For large enough $\delta \Omega_0$, stable hybrid configurations are not possible at all. The effect of increasing the coupling constant $g_v$ is very similar.  
We show that stable hybrid configurations with a maximum mass larger than the observed mass of the pulsar PSR J1614-2230 are possible for a large region of the parameter space of $g_v$ and $\delta \Omega_0$ provided the hadronic equation of state contains nucleons only. When the baryon octet is included in the hadronic phase, only a very small region of the parameter space allows to explain the mass of PSR J1614-2230. We compare our results with previous calculations of hybrid stars within the NJL model. 
We show that it is possible to obtain stable hybrid configurations also in the case $\delta \Omega_0=0$ that corresponds to the conventional NJL model for which the pressure and density vanish at zero temperature and chemical potential. 
\end{abstract}


\section{Introduction}
The recent determination of the mass of the pulsar PSR J1614-2230 with $1.97 \pm 0.04 M_\odot$ by \cite{Demorest},  renewed the discussions about the possibility of exotic matter being present at  the core of neutron stars. Since the description of matter at densities beyond nuclear saturation is model dependent, several works have explored different aspects of the fact that the maximum neutron star mass implied by any equation of state must exceed the mass of PSR J1614-2230 \citep{Weissenborn2011a,Weissenborn2011b,Weissenborn2012,Bednarek2011,Lastowiecki2011,Bonanno2012}. 
Some authors have revisited the role of hyperons in the equation of state showing that it is possible to construct stiff equations of state (EoS) with hyperons that are compatible with up-to-date hypernuclear data \citep{Bednarek2011}.  Others have investigated the role of the vector meson hyperon coupling, going from SU(6) quark model to a broader SU(3) symmetry \citep{Weissenborn2011b} and of hyperon potentials \citep{Weissenborn2012} in order to determine their impact on the maximum mass of neutron stars.

Concerning quark matter, it is known that models of strange stars made of absolutely stable quark matter satisfy comfortably the new constraint if color-superconductivity is taken into account \citep{Lugones2003,Horvath2004}.
However, it is not straightforward to construct models of hybrid stars with more than two solar masses \citep{Benhar2011,Weissenborn2011a}. A recurrent difficulty is that most hybrid EoS don't have at the same time a stable quark matter core and a sufficiently large maximum mass. For instance, most versions of the widely used Nambu-Jona-Lasinio (NJL) model are too soft to meet any of the above requirements. In a recent work,  \cite{Benhar2011}  performed a systematic study of the role of the vector and instanton-induced terms in the NJL Lagrangian and their connection with  the properties of hybrid stars. They explored a broad region of the parameter space showing that the instanton-induced interaction does not affect the stiffness of the quark matter EoS, whereas the effect of the repulsive vector interaction is significant. However, according to these authors, no values of the corresponding coupling constants allow for the formation of a stable core of quark matter \citep{Benhar2011}. These conclusions are in qualitative agreement with previous results using a similar EoS for the hadronic phase but including color superconductivity and neglecting the vector interaction term in the NJL EoS \citep{Baldo2003}, and with NJL models that implement the use of a density-dependent cutoff \citep{Baldo2007}.
Also, \cite{Jaziel} use a SU(2) NJL model with a vector term  but they still find unstable hybrid stars.
 Within a different picture, \cite{cesar2010} show that it is possible to obtain a stable sequence of compact stars with a quark core using a NJL model with an \textit{ad hoc} momentum cutoff that depends on the baryon chemical potential. However, within this approach, the maximum mass is still below the mass of PSR  J1614-2230. According to  \cite{Benhar2011},  their results are not essentially affected by the assumption that the hadronic phase consists of nucleons only,  or by the formation of mixed phases. However, more recent work by \cite{Bonanno2012} has succeeded in obtaining very massive stable hybrid configurations using a three-flavor NJL model that contains two free parameters: the transition density from hadronic matter to quark matter and the vector coupling of quarks. They show that high-mass stable configurations with color-superconducting quark cores can be constructed if vector interactions are included in the quark phase and if a very stiff hadronic equation of state is employed (e.g. the NL3 model with hyperons or the GM3 model with nucleons).

In this work  we perform an extensive study of hybrid star masses using several parametrizations of a relativistic mean-field hadronic EoS together with a typical three-flavor NJL model with scalar, vector and 't Hooft interactions. Within this approach, the hadronic and the quark-gluon degrees of freedom are derived from different Lagrangians and the deconfinement transition is associated with the point where both models have the same free energy. Thus, by construction, chiral symmetry restoration in the quark model occurs at a chemical potential $\mu$ that is in general different to the $\mu$ of deconfinement. Although this behaviour is an artefact of the hybrid EoS, it might be related with the actual properties of high density matter. Present numerical simulations of lattice QCD  indicate that for small chemical potential $\mu$, the deconfinement and chiral transitions coincide, and are crossover. At finite $\mu$, a chiral critical end point may exist in the phase diagram \citep{Stephanov2005}. It has been conjectured that, if such a critical end point exists,  the deconfinement and chiral transitions split from one another at that point \citep{McLerran2007}. As a consequence, the so called quarkionic phase (confined but chirally symmetric) could be present in the phase diagram \citep{McLerran2007}. In a similar way, we shall consider that the split between deconfinement and chiral symmetry restoration arising in the here-employed phenomenological hybrid EoS might represent an actual behaviour of matter at low temperatures and very high densities. In our model the unknown magnitude of this hypothetical split will be studied parametrically  by changing the conventional value of the vacuum pressure $-\Omega_0$ which is usually introduced in the grand thermodynamic potential of the NJL model in order to force a vanishing pressure at zero temperature and $\mu=0$. Instead of $-\Omega_0$ , we shall use a value $- (\Omega_0 + \delta \Omega_0)$, where $\delta \Omega_0$ is a free parameter. Since a change in the value of $\Omega_0$ has no influence on the fittings of the vacuum values for the pion mass, the pion decay constant, the kaon mass, the kaon decay constant, and the quark condensates, we shall treat it here as a free parameter. Analogously, the value of the vector coupling constant $g_v$ can be treated as a free parameter because the masses of the vector mesons are not dictated by chiral symmetry. 

The paper is organized as follows. In Sec. \ref{SecEoSH} we briefly present the here-employed hadronic EoS. In Sec. \ref{SecEoSQ} we describe the NJL SU(3) model  used to describe the quark matter paying particular attention to the role of the parameters $\delta \Omega_0$ and $g_v$. In Sec. \ref{SecResults} we present our results  and in Sec. \ref{Discussion}  we draw our main conclusions.

\section{Hadronic Phase}
\label{SecEoSH}

The relativistic mean-field  model is  widely used to describe hadronic matter in compact stars. 
In this paper we adopt the following standard Lagrangian \citep{Boguta1977,Glendenning2}: 
\begin{eqnarray} 
\label{baryon-lag}   
{\cal L}_H & = & \sum_{B} \bar{\psi}_{B}[\gamma_{\mu}(i\partial^{\mu}  - g_{\omega B}\omega^{\mu} - \frac{1}{2} g_{\rho B}\vec \tau . \vec \rho^{\mu})  \nonumber \\ 
& - & \left( m_{B} - g_{\sigma B}\sigma \right)]\psi_{B} + \frac{1}{2}({\partial_\mu \sigma \partial^\mu \sigma - m_{\sigma}^2 \sigma^2 } ) \nonumber \\ 
& - & \frac{1}{4} \omega_{\mu \nu}\omega^{\mu \nu}+ \frac{1}{2} m_{\omega}^2 \omega_\mu \omega^\mu - 
\frac{1}{4} \vec \rho_{\mu \nu}.\vec \rho^{\mu \nu} \nonumber \\
& + & \frac{1}{2} m_\rho^2 \vec \rho_{\mu}. \vec \rho^{\mu} -\frac{1}{3}bm_{n}(g_{\sigma}\sigma)^{3}-\frac{1}{4}c(g_{\sigma}\sigma)^{4}  \nonumber \\
& + & \sum_{L} \bar{\psi}_{L}    [ i \gamma_{\mu}  \partial^{\mu}  - m_{L} ]\psi_{L},
\label{eq1}
\end{eqnarray}
for matter composed  by (i) nucleons and electrons, and (ii) the baryon octet and electrons.   
Leptons $L$ are treated as non-interacting and baryons $B$ are coupled to the scalar meson $\sigma$, the isoscalar-vector meson $\omega_\mu$ and the isovector-vector meson $\rho_\mu$. For more details about the EoS obtained from the above Lagrangian the reader is referred to e.g. \cite{Lugones2010} and references therein. 
There are five constants in the model that are fitted to the bulk properties of nuclear matter  \citep{Glendenning2}.  In this work we use three different parametrizations shown in Table \ref{table1}.  
For all parametrizations we use a composition of nucleons and electrons. For the NL3 parametrization we consider also the case with the baryon octet and electrons. The parametrization for the hyperon coupling constants
is $g_{\omega \Lambda} / g_{\omega N} = g_{\omega \Sigma} / g_{\omega N} = 0.6666$,  $g_{\omega \Xi} / g_{\omega N}  = 0.3333$, $g_{\sigma \Lambda} / g_{\sigma N} = 0.6106 $,  $g_{\sigma \Sigma} / g_{\sigma N} = 0.4046$,  $g_{\sigma \Xi} / g_{\sigma N}  = 0.3195$ and   $g_{\rho i} / g_{\rho N}  = 1$  \citep{Chiapparini2009}. At low densities we use the Baym, Pethick and Sutherland (BPS) model \citep{BPS}. 
%
\begin{table}[t]
 \caption{Coupling constants for the parametrizations GM1 \citep{Glendenning2}, TM1 \citep{Sugahara1994} and NL3 \citep{Lalazissis}.  $M_{max}$  is the maximum mass of a pure hadronic star for matter composed by nucleons and electrons.  }
 \begin{center}
  \begin{tabular}{c|ccc} \hline \hline
      Set       &         GM1             &      TM1              &     NL3                \\ \hline

$m_\sigma$ (MeV)& 512                     &  511.198              &508.194               \\
$m_\omega$ (MeV)& 783                     &  783                  &782.501               \\
$m_\rho$ (MeV)  & 770                     &  770                  &763                   \\ \hline  
 $g_\sigma$     &\ \ 8.91 \ \             & \ \ 10.029  \ \       &  \ \ 10.217  \ \    \\
 $g_\omega$     &\ \   10.61 \ \          & \ \      12.614 \ \   &  \ \ 12.868      \ \   \\ 
 $g_\rho$       &\ \   8.196   \ \        &  \ \     9.264  \ \   &  \ \ 8.948  \ \    \\ 
  $b$           &  \ \ \ 0.002947 \ \ \   & \ \ \ -0.001506 \ \ \ & \ \  0.002055 \ \ \ \\
  $c$           &  \ \ \  -0.001070 \ \ \ & \ \ \  0.000061\ \ \ & \ \ \ -0.002651 \ \    \\
  $M_{max}$ \ \ &\ \   2.32      \ \      &  \ \     2.18    \ \  &  \ \   2.73   \ \              \\   
  \hline \hline
 \end{tabular}    
 \end{center}
 \label{table1}
\end{table}

\section{Quark Phase}
\label{SecEoSQ}

\subsection{The model} \label{njlmodel}

To describe the quark matter phase we use the SU(3) NJL model with scalar-pseudoscalar, isoscalar-vector and  't Hooft six fermion interaction. The Lagrangian density of the model is:
\begin{eqnarray}
{\cal L}_Q & = & \bar\psi(i \gamma_\mu \partial^\mu - \hat{m})\psi  \nonumber \\
           & + &  g_s \sum_{a=0}^{8} [( \bar\psi \lambda^a \psi )^2  +  (\bar \psi i \gamma_5 \lambda^a \psi)^2 ]  \nonumber \\
           & - &  g_v \sum_{a=0}^{8}   [(\bar\psi \gamma_\mu  \lambda^a \psi)^2 +  (\bar\psi \gamma_5 \gamma_\mu \lambda^a\ \psi)^2 ] \nonumber \\
           & + &  g_t\{\det[ \bar\psi(1+\gamma_5)\psi]   +  \det[ \bar\psi(1-\gamma_5)\psi]\}, 
\label{eq2}
\end{eqnarray}
where $\psi = (u,d,s)$ denotes the quark fields,  $\lambda^a ( 0 \leq a \leq 8 )$ are  the U(3) flavour matrices, 
 $\hat{m} = \mbox{diag}({m}_{u},{m}_{d},{m}_{s})$ is the quark current mass, and $g_s$, $g_v$ and $g_t$ are coupling constants.  

Notice that we have not included a diquark interaction term in the Lagrangian. As shown by  \cite{Ruster2005}, color superconducting phases (at $T=0$) are favoured in the regime of strong diquark coupling, $g_d /  g_s \approx 1$. However, in the regime of intermediate diquark coupling strength, $g_d / g_s = 3/4$, color superconductivity appears only above a chemical potential $\mu \sim 3 \times 440 ~ \mathrm{MeV} =  1320 ~ \mathrm{MeV}$ (see Fig. 1 of \cite{Ruster2005}).  For weaker diquark coupling, color superconductivity is shifted to very large densities that are not present at neutron star cores. Since the case of strong diquark coupling has already been considered by \cite{Bonanno2012} and \cite{Pagliara2008}, we shall focus here in a case where color superconductivity is negligible. 

The mean-field thermodynamic potential density $\Omega$ for a given baryon chemical potential $\mu$ at $T = 0$, is given by
\begin{eqnarray}
\Omega  = & - & \eta N_c\sum_i\int_{k_{Fi}}^{\Lambda}{\frac{p^2\, dp}{2\pi^2}}\sqrt{p^2+M_i^2} + 2g_s\sum_i \langle \bar \psi \psi \rangle_i^2  \nonumber \\
   & - & 2 g_v \sum_i \langle \psi^\dagger \psi \rangle_i^2 + 4g_t\langle\bar u u \rangle \langle\bar d d \rangle \langle\bar s s \rangle \nonumber  \\ 
  &  - & \eta N_c\sum_i \mu_{i}{\int_0^{k_{Fi}}  \frac{p^2\, dp}{2\pi^2}}-\Omega_0, 
\label{eq3}
\end{eqnarray}
where the sum is over the quark flavor $(i = u, d, s)$, the constants $\eta = 2$ and $N_c = 3$  are the spin and color 
degeneracies, and  $\Lambda$ is a regularization ultraviolet cutoff to avoid divergences in the medium integrals.
The Fermi moment of the particle $i$ is given by $k_{Fi} = \theta(\mu^{\ast}_{i} - M_i)\sqrt( \mu^{\ast 2}_{i} - M_i^2)$, where $\mu^\ast_{i}$ is the quark chemical potential modified by the vectorial interaction, i.e.
$\mu^{\ast}_{u,d,s} = \mu_{u,d,s} - 4 g_v \langle \psi^\dagger \psi \rangle_{u,d,s}$. 

In this work we consider the following set of parameters \citep{Kunihiro1989,Ruivo99}:  
$\Lambda = 631.4$ MeV , $g_s\Lambda^2 = 1.829$, $g_t\Lambda^5 = -9.4$, $m_u = m_d = 5.6$ MeV, $m_s = 135.6$ MeV, in order to fit the vacuum values for the pion mass, the pion decay constant, the kaon mass, the kaon decay constant, and the quark condensates: $m_\pi = 139.0$  MeV, $f_\pi = 93.0 $ MeV,  $m_K = 495.7 \mbox{ MeV}$, $f_K = 98.9 \mbox{ MeV}$, $\langle u \bar{u}\rangle^{1/3} = \langle u \bar{u}\rangle^{1/3} = -246.7 \mbox{ MeV}$, $\langle s \bar{s}\rangle^{1/3} = -266.9$ MeV.  The value of the vector coupling constant $g_v$ is treated a free parameter because the masses of the vector mesons are not dictated by  chiral symmetry. 
In order to obtain the equation of state, we assume that matter is charge neutral and in equilibrium under weak interactions.

\subsection{$\Omega_0$ as a free parameter}

\begin{figure}
\includegraphics[width = 0.45\textwidth]{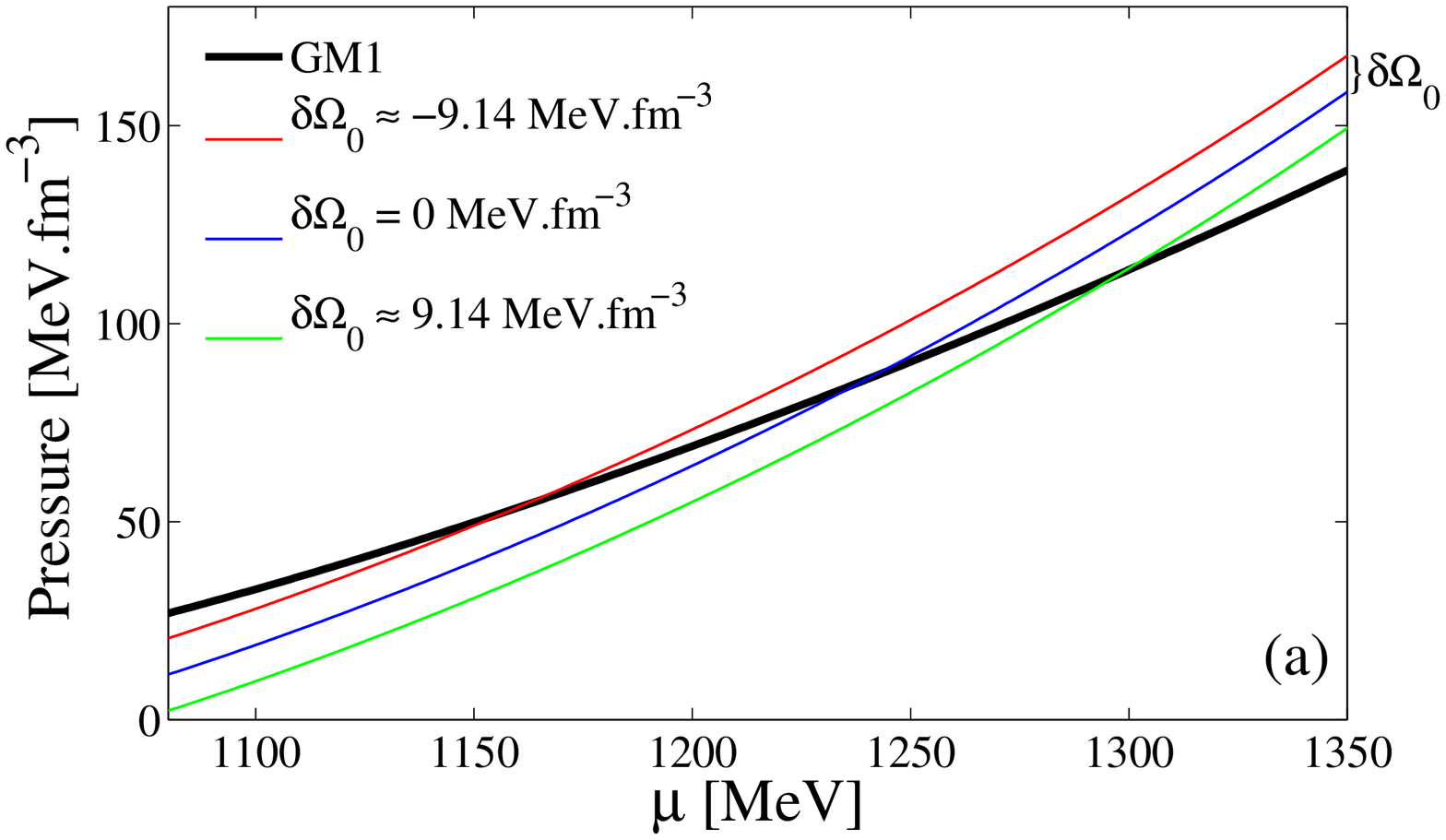}
\includegraphics[width = 0.45\textwidth]{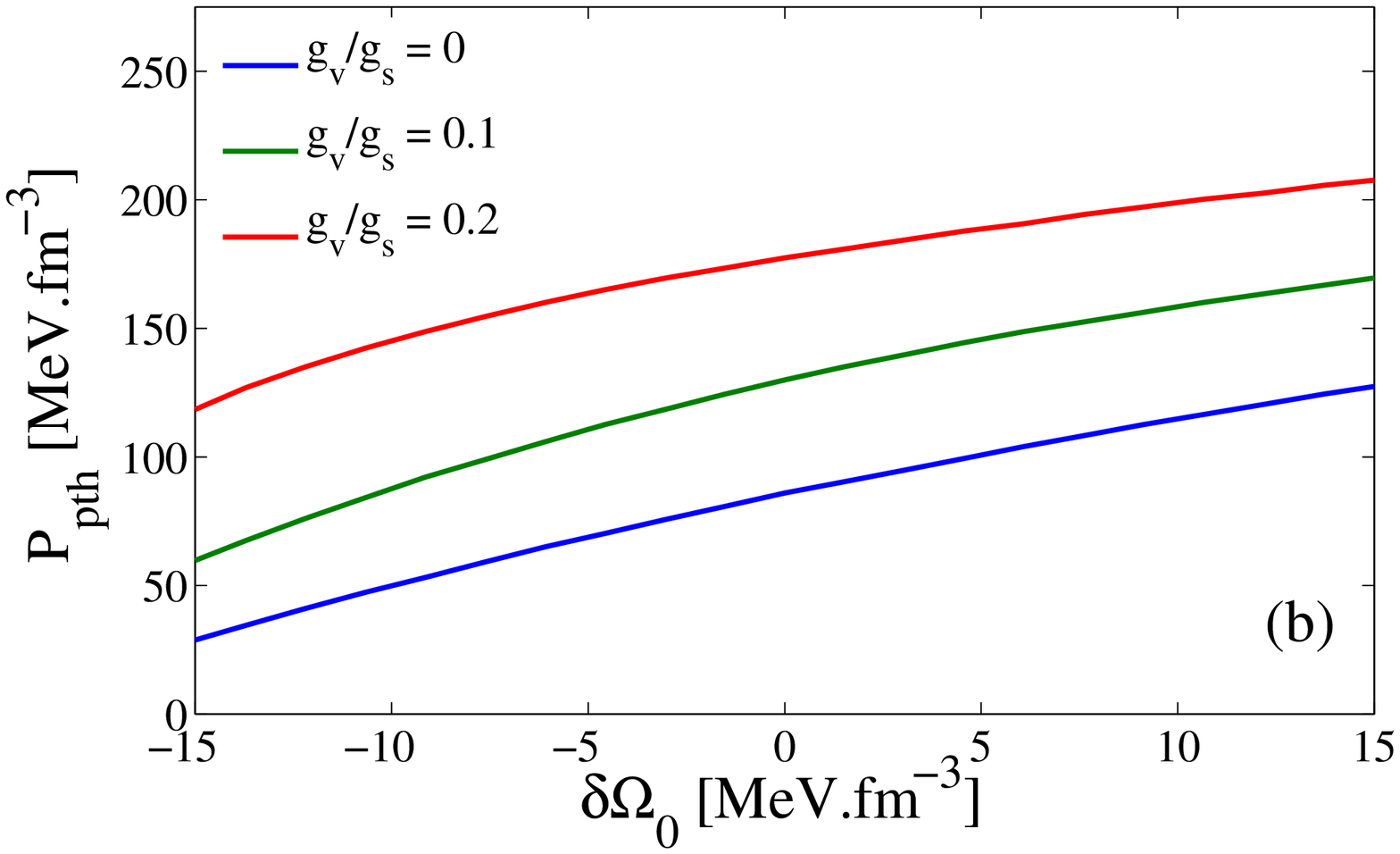}
\caption{(a) Pressure as a function of the chemical potential for different values of the parameter $\delta\Omega_0$. (b) Pressure of the deconfinement phase transition as a function of $\delta\Omega_0 $ for different values of the coupling constant $g_v$.  Notice that a small change in $\delta\Omega_0 $ can produce a significant  change in the pressure of the phase transition.}
\label{fig1}
\end{figure}

The conventional procedure for fixing the $\Omega_0$ term in Eq. (\ref{eq3}) is to assume that the grand thermodynamic potential $\Omega$ must vanish at zero $\mu$ and $T$. For the above quoted parametrization, this assumption leads to the value
$\Omega_0 = 5076.2 \ \mathrm{MeV \ fm}^{-3}$. Nevertheless, this prescription is no more than an arbitrary way to uniquely determine the EoS of the NJL model without any further assumptions \citep{Schertler1999}. Furthermore, in the MIT bag model for instance, the pressure in the vacuum is non-vanishing. In view of this,  \cite{Pagliara2008} adopt a different strategy.
They fix a bag constant for the hadron-quark deconfinement to occur at the same chemical potential as the chiral phase transition.
This method leads to a significant change in the EoS with respect to the conventional procedure. 

The connection between the chiral and the deconfinement phase transitions along the QCD phase diagram has received considerable attention in recent years (see \cite{Fukushima2011} and references therein). For zero chemical potential,  lattice results show that both transitions occur at the same temperature \citep{Karsch2002,Laermann}. At finite baryon chemical potential, this coincidence is an open question \citep{Fukushima2011}. However, since a chiral critical end point may exist in the plane of $T$ and $\mu$ \citep{Stephanov2005}, 
it has been conjectured  that the deconfinement and chiral transitions split from one another at that point \citep{McLerran2007}.  As a consequence,  a confined but chiral symmetric phase, which is called quarkyonic phase can exist in the high baryon density region. Since this conjecture is based on arguments that are valid in the large $N_c$ limit, it is not clear whether this quarkyonic phase can exist in the real QCD phase diagram. In the present paper we are not modelling the quarkionic matter because our confined phase is described by a hadronic model and the chiral transition is restricted to the quark phase. 
However, we may explore the above possibility of having chiral restoration and deconfinement occurring at different densities. To this end, we shall substitute $\Omega_0$ in Eq. (\ref{eq3}) by the new value $\Omega_0 + \delta \Omega_0$, where $\delta \Omega_0$ is a free parameter:
\begin{eqnarray}
\Omega_0  \longrightarrow  \Omega_0 + \delta \Omega_0  \  \ \ \ \   \mathrm{in} \ \mathrm{Eq.} \ (\ref{eq3}) .
\label{eq4}
\end{eqnarray}
With this change, the thermodynamic potential $\Omega$ can be non-vanishing at zero $\mu$ and $T$, and the $\mu$ of the deconfinement transition can be tuned. Clearly, $\delta \Omega_0$ has a minimum value because the phase  transition cannot be shifted to a pressure regime where the NJL model describes the vacuum. That is, we fix a minimum limit to $\delta \Omega_0$ for which the phase transition occurs at the chiral symmetry restoration point as performed by \cite{Pagliara2008}. On the other hand, there is no maximum value in principle for  $\delta \Omega_0$, since the phase transition can be shifted to arbitrarily large pressures. 

In order to illustrate the dependence of the EoS on the new parameter $\delta \Omega_0$ we depict in Fig. \ref{fig1}  the pressure as a function of the chemical potential for different values of $\delta\Omega_0$ and 
the pressure of the deconfinement transition $P_{pht}$ as a function of  $\delta \Omega_0$. Notice that a small change in the value of $\delta \Omega_0$ may result in a significant modification of the phase transition density, and consequently, in a very different hybrid EoS.

\section{Results}
\label{SecResults}

%
\begin{figure}
\centering
\includegraphics[width = 0.52\textwidth]{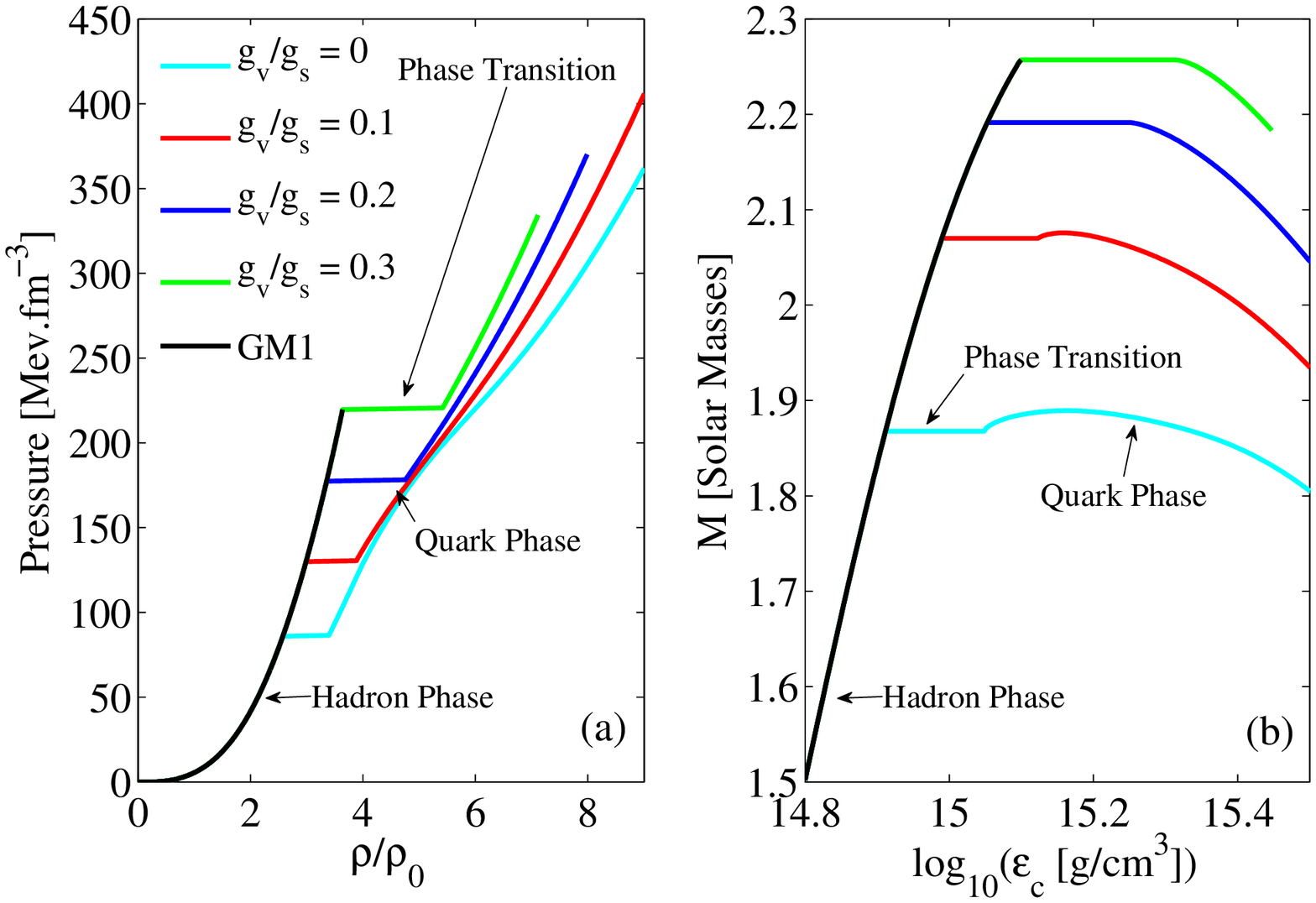}
\caption{(a) Pressure as a function of the baryon number density in units of the nuclear saturation density $\rho_0$ (we assumed $\rho_0 = 0.17 \mbox{ fm}^{-3}$).  (b) Mass
of  hybrid stars as a function of the central mass-energy density $\epsilon_c$. We use $\delta\Omega_0 = 0$ and different values of $g_v$.}
\label{fig2}
\end{figure}
%
\begin{figure}
\centering
\includegraphics[width = 0.52\textwidth]{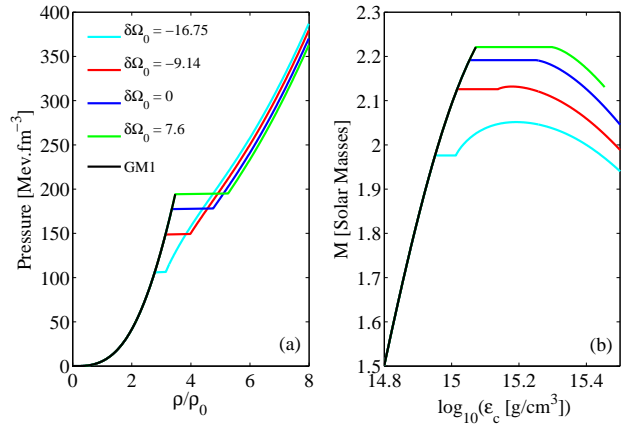}
\caption{Same as Fig. \ref{fig2} but adopting $g_v /g_s = 0.2$ and different values of $\delta\Omega_0$ (labels for $\delta\Omega_0$ are in $\mathrm{MeV \ fm}^{-3}$). }
\label{fig3}
\end{figure}
%
\begin{figure}
\includegraphics[width = 0.5 \textwidth]{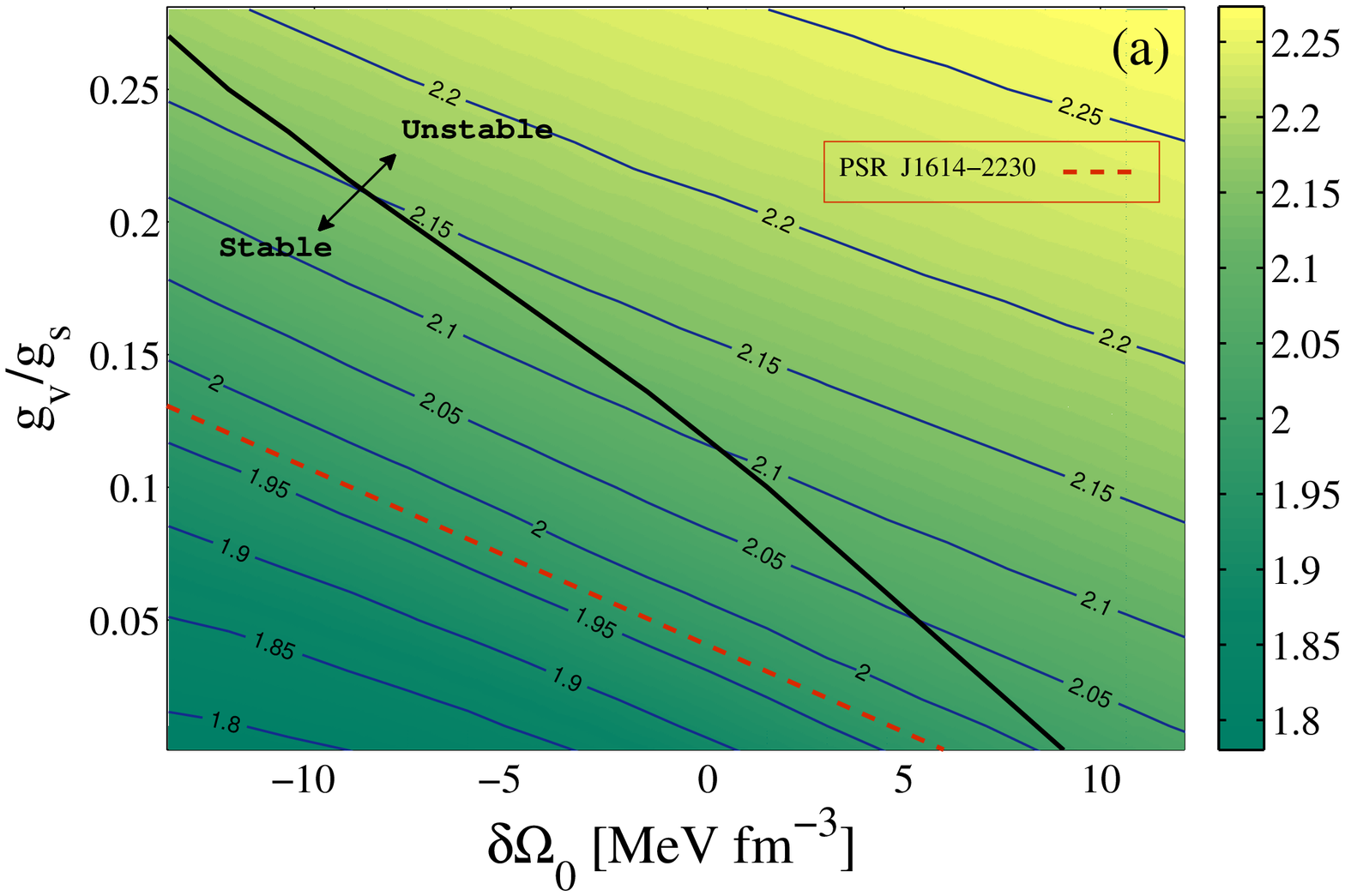}
\includegraphics[width = 0.5 \textwidth]{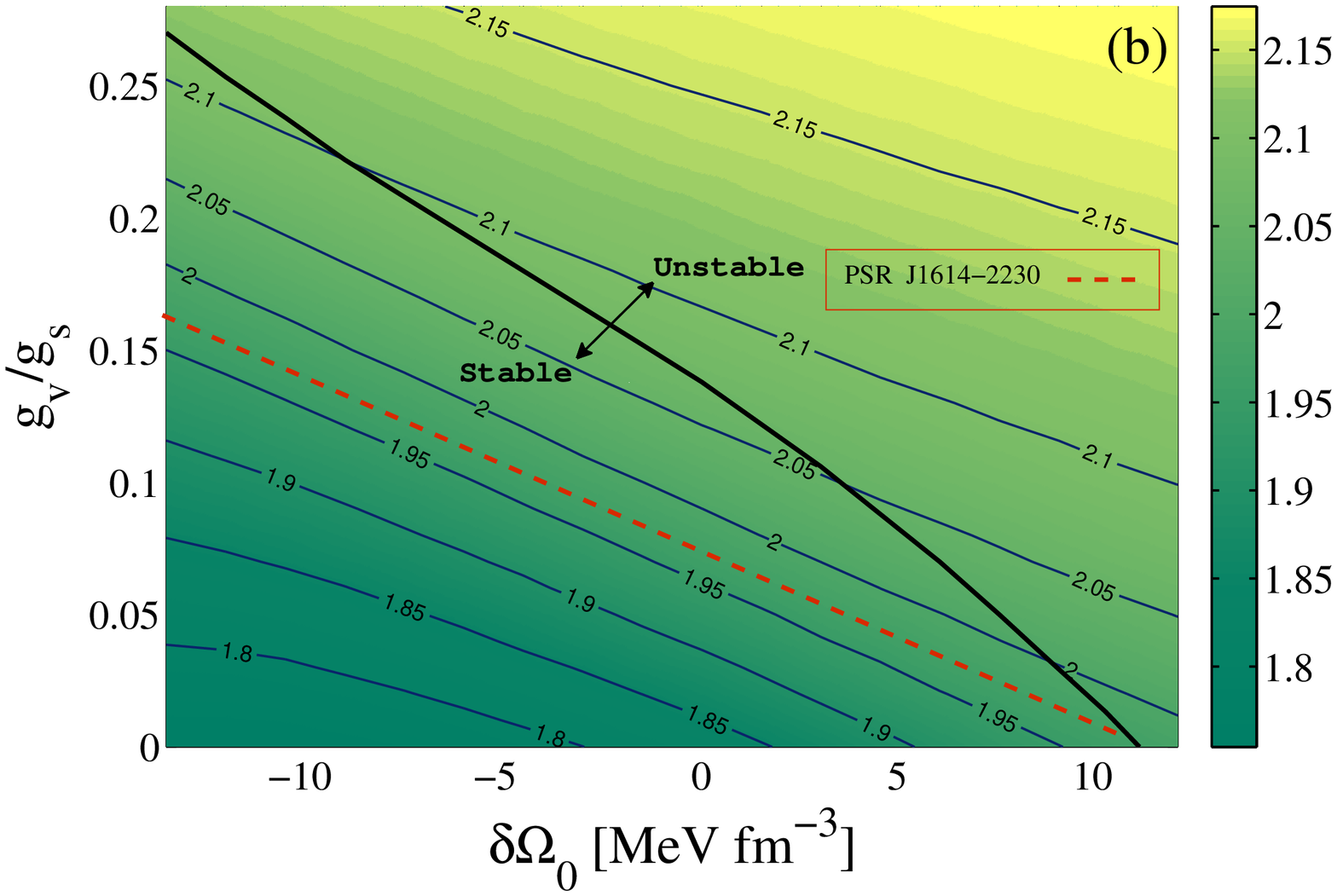}
\includegraphics[width = 0.5 \textwidth]{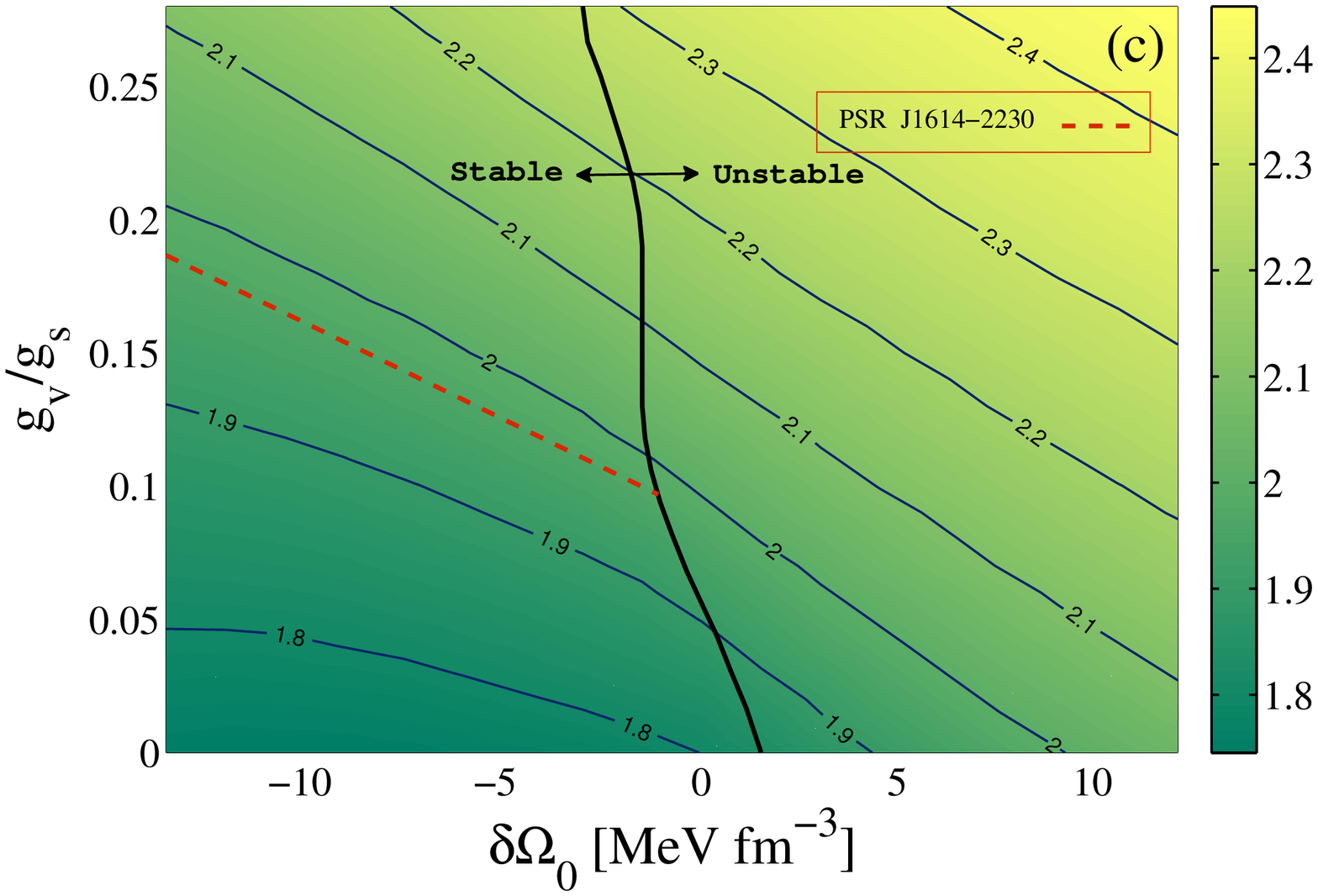}
\caption{Background colors represent the maximum mass of hybrid stars for different parametrizations of the NJL model (i.e. different values of $g_v$ and $\delta\Omega_0$).  In each panel we use a different hadronic EoS (without hyperons):  (a) GM1, (b) TM1 and (c) NL3 (see Table \ref{table1}). Notice that the color scale is different for each panel. The solid contour lines indicate specific values of the maximum mass. The black solid line represents the boundary between parametrizations that allow for stable hybrid stars and parametrizations that do not. The red dashed line indicates the value $1.97 {M}_{\odot}$  corresponding to the observed mass of PSR J1614-2230 \citep{Demorest}. The region between the red dashed line and the solid black line allows to explain the mass of PSR J1614-2230.}
\label{contour}
\end{figure}

We have solved the Tolman-Oppenheimer-Volkoff  equations for spherically symmetric and static stars in order to investigate the influence of $g_v$ and $\delta \Omega_0$ on the maximum mass of hybrid stars. 

In Figs. \ref{fig2} and \ref{fig3} we show the EoS for some specific parametrizations and the corresponding stellar configurations in a diagram of mass $M$ versus central energy density $\epsilon_c$. The plateaus represent the hadron-quark phase transition as a consequence of a first order Maxwell construction. 
In the left panel of Fig. \ref{fig2} we note that as we increase the value of the vector coupling constant $g_v$ so does the density of the phase transition, and therefore, the hybrid star has a smaller quark core and a larger hadronic contribution. This leads to larger maximum masses because the hadronic EoS is stiffer than the quark EoS. At the same time, there is a larger density jump between the two phases, which tends to destabilize the configuration.       
Due to these two effects, together with the fact that the vector term stiffens the NJL EoS, models with a larger $g_v$ give larger maximum masses but have stable quark cores within a smaller range of central densities (see right panel of Fig. \ref{fig2}). 
In the left panel of Fig. \ref{fig3} we show the effect of changing the magnitude of the shift between the deconfinement and the chiral phase transitions. As we increase $\delta\Omega_0$ from negative to positive values, we increase the density of the phase transition as well as the density jump between the two phases. However, the NJL EoS becomes slightly softer because there is a larger contribution to the EoS of the regime with a partially restored chiral symmetry. Since the latter effect is not so strong, the impact of increasing $\delta\Omega_0$ is analogous to increase $g_v$, i.e. models with a larger $\delta\Omega_0$ result in larger maximum masses but the quark cores are stable within a smaller range of central densities (see right panel of Fig. \ref{fig3}).  

In Fig. \ref{contour}, background colors represent the maximum mass of hybrid stars for different values of $g_v$ and $\delta\Omega_0$.  We used the hadronic EoS of Table \ref{table1} with nucleons and electrons. Within each panel we show contour lines indicating specific values of the maximum mass. The black line represents the limit between parameters that allow for stable hybrid stars and those that always give unstable hybrid stars. The value $1.97  {M}_{\odot}$ corresponding to the observed mass of PSR J1614-2230  \citep{Demorest} is shown with a red dashed line. An interesting feature of Fig. \ref{contour}, is that large masses are situated on the right-upper corner but stable configurations are located on the left-lower corner of the figure (or left side of the figure in the case of NL3). This clearly illustrates the difficulty of obtaining stable hybrid stars with arbitrarily large masses. 
Concerning the effect of the hadronic model we see that stable hybrid stars have higher values of the maximum mass for the stiffer hadronic EoS. 

The observed mass of PSR J1614-2230 can be explained by parameters within the large region located between the red dashed line and the solid black line in each panel of Fig. \ref{contour}.  However, a hypothetical  future observation of a neutron star  with a mass a  $\sim 10 \%$ larger than the mass of PSR J1614-2230 will be hard to explain within hybrid star models using the GM1 and TM1 EoS (see panels (a) and (b) of Fig. \ref{contour}) and will require a very stiff hadronic model such as NL3.

\begin{figure}
\begin{center}
\includegraphics[width = 0.4 \textwidth]{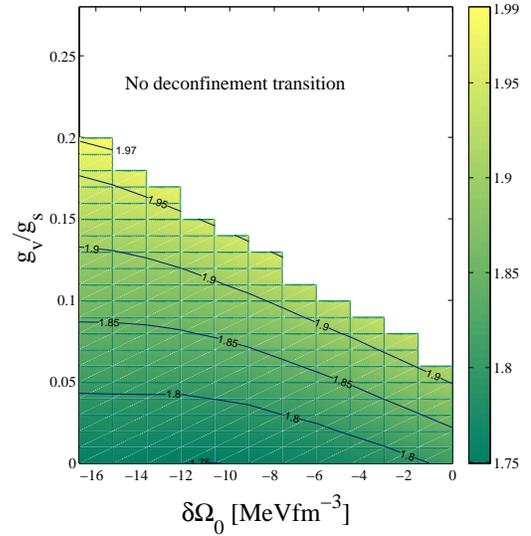}
\end{center}
\caption{Same as panel (c) of Fig. 4 but for the NL3 model with hyperons. Hybrid stars are not possible for the set of parameters within the white region. Only a very small region near the upper-left corner of the coloured region  allows to explain the mass of PSR J1614-2230.}
\label{fig5}
\end{figure}

The effect of hyperons is shown in Fig. \ref{fig5} where we consider the  NL3 parametrization with the inclusion of the baryon octet. Compared with the case without hyperons, the maximum mass values are altered by a few percent. This follows from the fact that the deconfinement phase transition occurs at relatively low densities, i.e. in regions where the baryon octet has a minor contribution. Nevertheless, hyperons have a large effect in the possibility of finding stable hybrid stars of large enough mass. When we increase the value of $g_v$  or increase  $\delta \Omega_0$, the deconfinement transition is shifted to larger densities and  the hadronic EoS with hyperons tends to be favoured almost everywhere in the star. Above a certain limit there is no deconfinement transition at all, i.e. the star is always hadronic (see white region in Fig. 5). As a consequence, the mass of PSR J1614-2230 can be attained for models within a very small region of the parameter space located near the upper-left corner of the coloured region in Fig. 5.

\section{Discussion and Conclusions}
\label{Discussion}

In this work we performed a systematic study of hybrid star configurations using a relativistic mean-field hadronic EoS and the NJL model for three-flavor quark matter. For the hadronic phase we used the stiff GM1 and TM1 parametrizations, as well as the very stiff NL3 model. In the  NJL Lagrangian we included scalar, vector and 't Hooft interactions. 
The vector coupling constant $g_v$ was treated as a free parameter.  We also considered that there is an arbitrary split between the deconfinement and the chiral phase transitions. This split can be adjusted by a redefinition of the constant parameter  $\Omega_0$  in the NJL thermodynamic potential; i.e. by making the replacement $\Omega_0  \longrightarrow  \Omega_0 + \delta \Omega_0$ in Eq. (\ref{eq3}), where $\delta \Omega_0$ a free parameter.  We find that, as we increase the value of $\delta \Omega_0$,  hybrid stars have a larger maximum mass but are less stable (i.e. hybrid configurations are stable within a smaller range of central densities). For large enough $\delta \Omega_0$, stable hybrid configurations are not possible at all (see Fig. 3). The effect of increasing the coupling constant $g_v$ is very similar (see Fig. 2).  
These effects are clear in Fig. \ref{contour},  where we show the maximum mass of static spherically symmetric stars in the parameter space of $g_v$ and $\delta \Omega_0$.  The larger masses are situated on the right-upper corner of the diagram, where both $g_v$ and $\delta \Omega_0$ are larger. On the other hand, stable configurations are placed on the opposite part of the diagram; i.e. on the left-lower corner for the GM1 and TM1 models and on the left side for the NL3 model.  As a consequence, stable configurations with a maximum mass compatible with PSR J1614-2230 are located halfway, specifically, between the red dashed line and the solid black line of Fig. 4. The effect of the hadronic model (with nucleons only) is also clear from Fig. 4, where we see that stable hybrid stars have higher values of the maximum mass for the stiffer hadronic EoS. The main effect of hyperons is that they preclude the deconfinement transition in the region of the parameter space that allows large maximum masses (see Fig. 5). Only a very small area near the upper-left corner of the coloured region of Fig. 5 allows to explain the mass of PSR J1614-2230.

It is worth summarizing the main assumptions and results of recent work using the NJL model to describe hybrid stars. \cite{Schertler1999}  use several parametrizations of an extended relativistic mean field  model to describe the hadronic phase with hyperons. For the deconfined quark phase they use the NJL model with three flavors, including scalar and 't Hooft terms, and using the RHK parameter set \citep{Rehberg1996}. Their conclusion is that typical neutron stars with masses around 1.4 solar masses do not possess any deconfined quark matter in their center. 
More recently, \cite{Baldo2007} use a SU(2) NJL model with the RHK parametrization and a cut-off that depends on the chemical potential. For the hadronic phase they  adopt a nucleonic equation of state obtained within the Brueckner-Bethe-Goldstone approach using the Argonne $v_{18}$ two-body potential, supplemented by the Urbana phenomenological three-body force. They are not able to obtain stable hybrid stars. \cite{Jaziel} use a SU(2) NJL model with the RHK parametrization but adding a vector term together with the GM1 parametrization of a relativistic mean field hadronic model \citep{Glendenning2}. The density of the phase transition increases with  $g_v$ but  they still find unstable hybrid stars.  \cite{Benhar2011} work with SU(3) and the RHK parametrization. They use several values of $g_v$ and they also vary the coupling constant of the 't Hooft term. For the hadronic phase they use  a phenomenological Hamiltonian including the Argonne $v^{\prime}_6$ nucleon-nucleon potential. They conclude that no values of the corresponding coupling constants allow for the formation of a stable core of quark matter.  

All the above versions of the NJL model use the conventional procedure of imposing that the pressure and density must vanish at zero temperature and chemical potential. However, a different prescription is used by \cite{Pagliara2008}. First, they introduce a hadronic EoS and  compute the transition to quark matter by a Maxwell construction. To fix the bag constant they assume that deconfinement occurs at the same chemical potential as the chiral phase transition, i.e. they require that the pressure of quark matter is equal to the pressure of the hadronic matter at the critical chemical potential for which chiral symmetry is restored. The bag value obtained with this assumption is the lowest possible value for the bag constant in the NJL model because it allows to use the NJL EoS just starting from the chemical potential of the chiral phase transition. Using this procedure, \cite{Pagliara2008} computed the equation of state of quark matter within the NJL model by including effects from the chiral condensates, the diquark coupling pattern, and a repulsion vector term. They find that hybrid stars containing a CFL core are stable but the maximum mass is $\sim 1.8 M_{\odot}$, i.e. incompatible with PSR J1614-2230. More recently,  \cite{Bonanno2012} use a NJL model supplemented by the 't Hooft and vector interactions and consider the 2SC and CFL color superconducting phases. For the hadronic phase they use a relativistic mean field model and adopt the NL3 parameterization with hyperons and the GM3 parameterization with nucleons only. In both cases they are able to obtain  maximum masses above the mass of PSR J1614-2230 because they use the non-conventional procedure of treating the density of the deconfinement transition as a free parameter.

Our analysis is related to that of \cite{Bonanno2012} because we can control the density of the phase transition via the parameter $\delta \Omega_0$ in the thermodynamic potential. However, since our parameter space is constructed in terms of $\delta \Omega_0$ and not in terms of the deconfinement density as in \cite{Bonanno2012},  the connection with conventional NJL models is more transparent in our case. Additionally, our procedure includes the non-conventional prescription of  \cite{Pagliara2008} as a special case. 
It is also worth noticing that we use the parametrization of the NJL EoS given by \cite{Kunihiro1989,Ruivo99}
while the above authors use the RKH one which is somewhat softer (see \cite{Buballa2004} for more details on the parametrizations). Another difference is that they consider quark matter in the 2SC and CFL phases,  whereas we don't consider color superconductivity in the quark phase. In this sense, these works are complimentary because \cite{Pagliara2008,Bonanno2012} work in the regime of strong diquark coupling  ($g_d /  g_s \approx 1$) where color superconductivity is strongly favoured, whereas we work in the weak  diquark coupling regime ($g_d /  g_s \rightarrow 0$) for which color superconductivity is shifted to very large densities that are not present at neutron star cores (see \cite{Ruster2005} for more details).

In summary, our results show that hybrid configurations with maximum masses equal or larger than the observed mass of PSR J1614-2230 are possible for a significant region of the parameter space of $g_v$ and  $\delta \Omega_0$ provided a stiff enough hadronic EoS without hyperons is used. It is also worth highlighting the fact that we can obtain compact stars with stable quark cores without having to perform any modification to the NJL model (i.e. setting $\delta \Omega_0 = 0$). This is in contrast with the results obtained by \cite{Baldo2007}, \cite{Benhar2011} and \cite{Jaziel} who are unable to obtain stable hybrid star configurations with the NJL model. The difference arises because these authors use the softer parametrization of \cite{Rehberg1996} while we use the stiffer parametrization of \cite{Kunihiro1989} and \cite{Ruivo99}; i.e.  the use of  the latter parametrization allows to reproduce the mass of the pulsar PSR J1614-2230 without using  exceptionally stiff parametrizations of the hadronic EoS and keeping the conventional procedure for fixing $\Omega_0$. It is also interesting to note that we have typically  $\delta \Omega_0 /  \Omega_0 \sim 0.1 \%$, i.e. small departures of the conventional $\Omega_0$ are sufficient 
to obtain massive enough stable hybrid stars. Finally, we emphasize that the observation of compact star masses a few percent larger than the mass of PSR J1614-2230 will be hard to explain within hybrid star models using the GM1 and TM1 EoS  and will require a very stiff hadronic model such as NL3 with nucleons only.

\begin{acknowledgements}
We acknowledge the financial support received from FAPESP-Brazil. We also acknowledge C. Provid\^encia and P.  Costa for helpful discussions.
\end{acknowledgements}

\end{document}